\documentclass[journal]{IEEEtran}

\usepackage[english]{babel}
\usepackage[noadjust]{cite}
\usepackage{algorithm}
\usepackage{algpseudocode}
\usepackage{optidef}
\usepackage[acronym]{glossaries}
\usepackage{makecell}
\usepackage{soul}
\ifCLASSINFOpdf
\else
   \usepackage[dvips]{graphicx}
\fi
\usepackage{url}
\usepackage{multirow}
\usepackage{tabularx}
\usepackage{graphicx}
\usepackage{amsmath}
\usepackage{bm}
\usepackage{amssymb}
\usepackage{xcolor}
\usepackage{amsthm}
\usepackage{booktabs}
\usepackage{multirow}
\usepackage{siunitx}

\newacronym{aiot}{A-IoT}{Ambient IoT}
\newacronym{3gpp}{3GPP}{3rd Generation Partnership Project}
\newacronym{iot}{IoT}{Internet of Things}
\newacronym{rfid}{RFID}{Radio Frequency Identification}
\newacronym{lpwan}{LPWAN}{low power wide area network}
\newacronym{tr}{TR}{technical report}
\newacronym{bs}{BS}{base station}
\newacronym{ue}{UE}{user equipment}
\newacronym{iab}{IAB}{integrated access and backhaul}
\newacronym{vwc}{VWC}{volumetric water content}
\newacronym{ook}{OOK}{on-off keying}
\newacronym{mimo}{MIMO}{multiple-input multiple-output}
\newacronym{csi}{CSI}{channel state information}
\newacronym{am}{AM}{amplitude modulation}
\newacronym{ask}{ASK}{amplitude-shift keying}
\newacronym{psk}{PSK}{phase-shift keying}
\newacronym{pm}{PM}{phase modulation}
\newacronym{pru}{PRU}{Positioning Reference Unit}
\newacronym{lmf}{LMF}{Location Management Function}
\newacronym{aoa}{AoA}{angle of arrival}
\newacronym{aod}{AoD}{angle of departure}
\newacronym{toa}{ToA}{time of arrival}
\newacronym{tdoa}{TDoA}{time difference of arrival}
\newacronym{fpga}{FPGA}{field-programmable gate array}
\newacronym{mcu}{MCU}{microcontroller unit}
\newacronym{rf}{RF}{radio frequency}
\newacronym{nbiot}{NB-IoT}{Narrowband IoT}
\newacronym{emtc}{eMTC}{enhanced machine-type communication}
\newacronym{ble}{BLE}{Bluetooth Low Energy}
\newacronym{bl}{BL}{battery-less}
\newacronym{ba}{BA}{battery-assisted}
\newacronym{bsa}{BSA}{battery and signal-assisted}
\newacronym{ag2ug}{AG2UG}{aboveground-to-underground}
\newacronym{ug2ag}{UG2AG}{underground-to-aboveground}
\newacronym{ml}{ML}{machine learning}
\newacronym{swipt}{SWIPT}{simultaneous wireless information and power transfer}
\newacronym{rrc}{RRC}{Radio Resource Control}

\begin{document}

\title{Ambient IoT: Communications Enabling Precision Agriculture}

\author{Ashwin Natraj Arun, Byunghyun Lee, Fabio A. Castiblanco, Dennis R. Buckmaster, Chih-Chun Wang, David J. Love,  James V. Krogmeier, M. Majid Butt and Amitava Ghosh.

\thanks{This work is supported in part by the National Science Foundation (NSF) under grants CNS2212565, EEC1941529, CCF2008527, CNS2107363, CNS2235134, CNS2225578, CCF2309887 and Nokia.
}
\thanks{Ashwin Natraj Arun, Byunghyun Lee, Fabio A. Castiblanco, Chih-Chun Wang, David J. Love and James V. Krogmeier are with the Elmore Family School of Electrical and Computer Engineering, Purdue University, West Lafayette, IN 47907, USA (e-mails: \{ashwin97, lee4093, fcastibl, chihw, djlove, jvk\}@purdue.edu).}
\thanks{Dennis R. Buckmaster is with the School of Agricultural and Biological Engineering, Purdue University, West Lafayette, IN 47907, USA (e-mail: dbuckmas@purdue.edu)}
\thanks{M. Majid Butt and Amitava Ghosh are with Nokia Standards, USA (e-mails: \{majid.butt, amitava.ghosh\}@nokia.com).}}
\maketitle

\IEEEpeerreviewmaketitle

\begin{abstract}
One of the most intriguing 6G vertical markets is precision agriculture, where communications, sensing, control, and robotics technologies are used to improve agricultural outputs and decrease environmental impact. 
\gls{aiot}, which uses a network of devices that harvest ambient energy to enable communications, is expected to play an important role in agricultural use cases due to its low costs, simplicity, and battery-free (or battery-assisted) operation.
In this paper, we review the use cases of precision agriculture and discuss the challenges. 
We discuss how \gls{aiot} can be used for precision agriculture and compare it with other ambient energy source technologies. We also discuss research directions related to both \gls{aiot} and precision agriculture. 
%
\end{abstract}

\glsresetall

\section{Introduction}
In recent decades, agriculture has undergone a paradigm shift driven by technological advancements to meet the demands of a growing population.
This evolution has given rise to precision agriculture, an innovation that leverages cutting-edge technologies to optimize resource utilization, enhance crop yields, and mitigate environmental impacts.
Precision agriculture aims to integrate data-driven decision-making, automation, and advanced sensor technologies to create a more efficient and sustainable agricultural system.

Concurrently, many economic sectors have been significantly changed by the rapid growth of \gls{iot} devices which offer simplification and lower costs in a wide range of applications.
This growth is expected to continue, with potential deployments in agriculture, healthcare and manufacturing markets \cite{6gverticalroadmaap}. In agriculture, the advent of IoT has enabled improved
systems for sensing data, communicating, inferring information and making decisions. 
Currently, standards such as \gls{nbiot} and \gls{emtc}, which are based on the \gls{3gpp}, as well as LoRaWAN, facilitate communication for IoT devices.
However, supporting a high density of IoT devices in rural areas remains challenging with the current wireless infrastructure \cite{zhang_rural_wireless}.
Precision agriculture relies on sensor data, GPS-guided machinery, and variable rate technology for optimal crop management. 
This approach enhances productivity, reduces resource waste, and supports sustainable farming through data-driven decisions. However, achieving this promise requires sensor deployment at high densities and commonly available communications, such as Wi-Fi, are inadequate for vast farm areas.
Instead, energy-harvesting devices using ambient sources for power and communication are needed to meet these requirements.
Devices that harvest energy from ambient sources such as electromagnetic, solar, and thermal to power and communicate and operate within the \gls{3gpp} network are known as \gls{aiot} devices.
\gls{aiot} devices can operate with or without a battery and do not require a constant power supply, relying on backscattering to communicate.

Backscatter communications have been around for decades and are used in technologies such as \gls{rfid} \cite{intro_rfid} and Wi-Fi backscatter \cite{ambient_back_comm_survey}.
However, there is little prior work on dense deployments and operations within the existing cellular architecture. 
Recently, 3GPP has started discussion on \gls{aiot} devices with potential use cases, relevant communication scenarios and topologies of operation for such devices. 
While there is research looking at the performance of \gls{aiot} devices in conventional use cases, there is very little literature on the application of \gls{aiot} devices in the domain on precision agriculture. 
We review the feasibility of \gls{aiot} devices for precision agriculture with a link-budget analysis and compare \gls{aiot} to \gls{rfid} devices. 
Also, we look at the challenges and potential research directions for \gls{aiot} devices. 
In this work, we aim to answer several fundamental questions relating precision agriculture and \gls{aiot}:
\begin{itemize}
    \item What are different use cases of \gls{aiot} in precision agriculture?
    \item What is backscattering and how does \gls{aiot} interface with existing \gls{3gpp} architecture?
    \item What are the challenges and potential research directions for \gls{aiot} in precision agriculture?
\end{itemize}

\begin{figure*}[!t]
    \centering
    \includegraphics[width=0.8\linewidth]{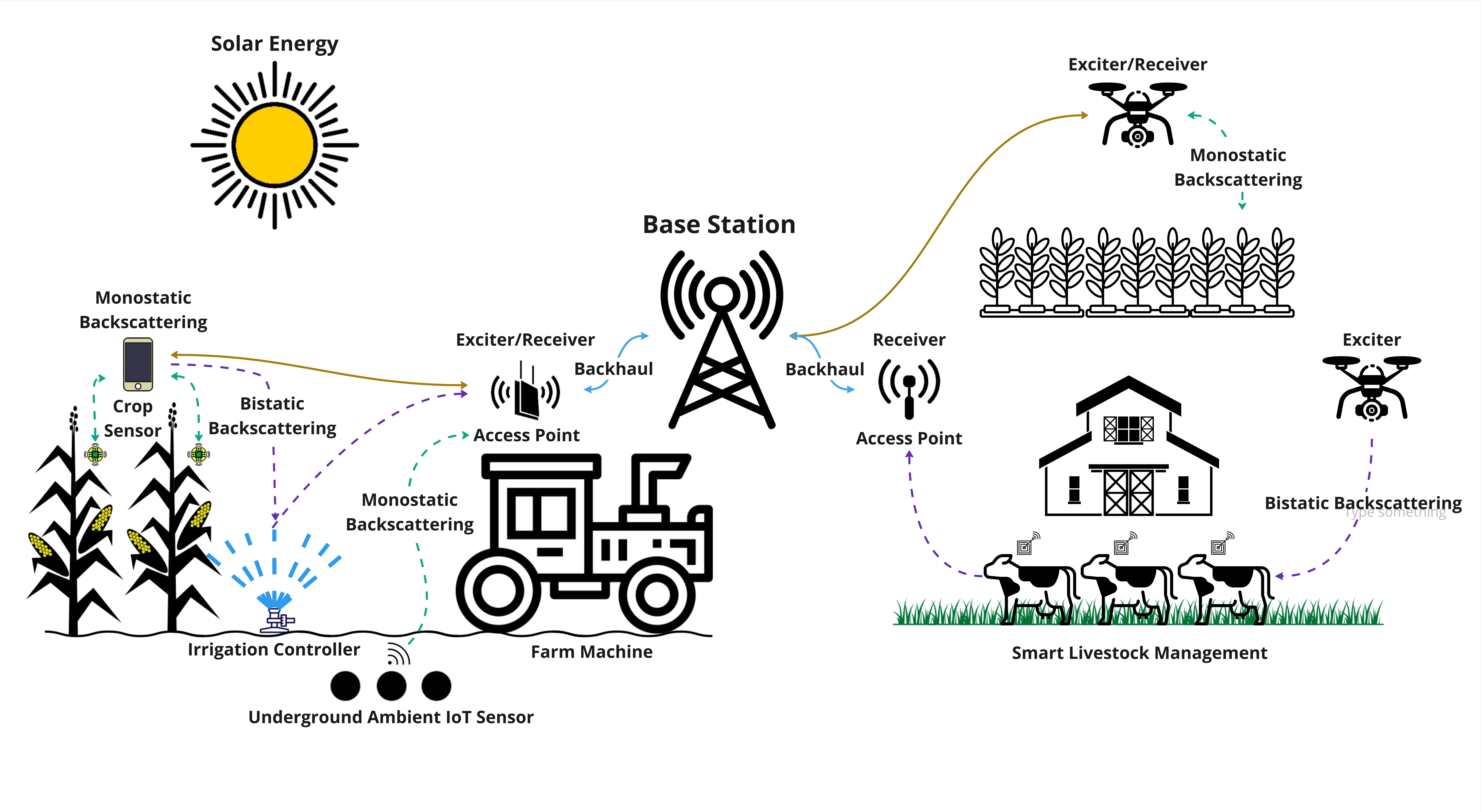}
    \vspace{-0.8cm}
    \caption{Landscape of \gls{aiot} devices used in Precision Agriculture.}
    \label{fig:A-IoT_landscape}
    \vspace{-0.5cm}
\end{figure*}
\vspace{-0.4cm}
\section{Precision Agriculture with \glsdesc{aiot}}\label{use_cases}
%
Precision agriculture uses sensors, GPS, drones, and other technologies to optimize crop management to enhance efficiency, increase yields, and minimize environmental impact.
Fig. \ref{fig:A-IoT_landscape} illustrates a landscape of use cases where precision agriculture benefits significantly from \gls{aiot} devices.
\vspace{-0.45cm}
\subsection{Sensing in Precision Agriculture}
\gls{aiot} devices can be used in precision agriculture to monitor environmental conditions in farms and greenhouses. 
These devices can measure parameters such as air temperature, humidity, $\text{CO}_2$ concentration, light, soil temperature, and pH. 
By monitoring these parameters, predictions can be made about soil health, crop yield, pest management etc. 
\gls{aiot} devices are ideal for this purpose because they do not require an external power source and are compact. 
This allows farmers to deploy a large number of sensors in fields to monitor both crops and soil efficiently. 
Sensors can be placed in open fields, greenhouses with \gls{aiot} devices connected to the network through various discussed topologies. 

\subsection{Actuation and Controllers in Precision Agriculture}
In agricultural production, various factors such as soil, climate, and water significantly affect crop growth and yield.
Agricultural equipment, like pesticide sprayers, fertilizer spreaders and irrigation systems, can be actuated or controlled periodically based on sensor data to ensure optimal yield, and effective crop management.
Providing a continuous power supply for these controllers in outdoor farms is a challenge. \gls{aiot} enables seamless communication for these control and actuation devices by harvesting ambient energy and triggering the controllers. 
This ensures reliable data transmission and remote control capabilities, even in areas with limited infrastructure. 
Since the controllers operate periodically, the \gls{aiot} precision agricultural controllers can be activated periodically by the farm management system. This integration of \gls{aiot} with the cellular architecture enhances the responsiveness and efficiency of agricultural operations.

\subsection{Underground Soil Sensing in Precision Agriculture}
Existing sensors used in precision agriculture are mostly above ground or have antennas above ground. 
To accurately measure nitrogen levels, soil moisture, and other important metrics, sensors need to be placed underground. 
Given the need for dense sensor deployment, it is essential to use devices that are small, low-cost, power-efficient, and biodegradable. 
\gls{aiot} devices are ideal for this use case due to these characteristics. 
Communication from \gls{aiot} devices to the network can be facilitated using a mobile relay node on farm machinery such as tractors, sprayers and harvesters. 
As these machines move across the field, the \gls{aiot} devices become active and send data to the node mounted on the machine.

In corn fields, there is a notable distinction between surface soil moisture and the moisture content deeper underground. 
Corn plants develop deep roots to access water, emphasizing the need for soil moisture monitoring at multiple depths.
Unlike conventional sensors that require periodic removal for data retrieval, underground \gls{aiot} devices offer the advantage of data collection without physical intervention, which is particularly beneficial for crops like corn.

\subsection{Smart Livestock Management}
Smart livestock farming employs innovative production systems to enhance sustainability and reduce food waste and crises. 
Body temperature is a vital health indicator for livestock and is crucial for farmers to identify and act on potential diseases early. 
Changes in body temperature signify illness, making it a precision health indicator. Importantly, acquiring body temperature data is not latency-intensive, and the increasing herd size necessitates the use of cost-effective ear tags over high-power IoT devices. 
\gls{aiot} devices are ideal for this application due to their low cost, small form factor and battery-free operation. 
Cattle and pigs can be equipped with small ear tags that monitor their body temperature and transmit data to the farm.

\subsection{Food Supply Chain}
Innovative solutions are essential to tackle food waste and ensure food safety. 
A controlled environment is crucial for fresh foods, such as vegetables and meat, to maintain their safety and shelf life. 
An efficient method is needed to monitor every stage of the food supply chain to reduce food waste. 
The increasing demand for organically sourced food items also requires a reliable way to guarantee production and processing methods in the food supply chain.
Equipping the food supply chain with \gls{aiot} devices in each transport item can address these needs. 
These devices update relevant information at each stage, from the seed stage to harvesting and packaging facilities. 
The end user can verify the produce by accessing the data from the \gls{aiot} device. Similarly, suppliers can use these devices to track real-time demand and stock accordingly, reducing food waste. 
This principle applies at each stage of the food supply chain, decreasing overstocking and waste.

\section{Backscatter Communications}
A-IoT devices use backscatter communication, a wireless technique where devices reflect and modulate an incoming \gls{rf} signal to transmit data \cite{xu2018practical}. 
This method is particularly useful for devices with limited or zero internal power, as it does not require the generation of new RF signals. 
There are two types of backscatter communication: monostatic and bistatic as seen in Fig. \ref{fig:A-IoT_landscape}.

In a monostatic system, the energy exciter and signal receiver are the same device, whereas in a bistatic system, they are separate devices. 
At the heart of backscatter communication lies impedance mismatching. 
Varying the load impedance of an \gls{aiot} device changes its reflection coefficient, affecting the amplitude and phase of the reflected wave. 
This allows the \gls{aiot} device to perform various modulation techniques, such as \gls{ask}, \gls{am}, \gls{psk}, \gls{pm}, and their combinations via load modulation \cite{xu2018practical}. 
The modulation order is proportional to the number of load impedance states in the \gls{aiot} device. 

Backscatter communication is widely used in technologies such as \gls{rfid}, Wi-Fi backscatter, and \gls{ble} backscatter. 
In \gls{rfid}, a dedicated \gls{rf} transceiver emits \gls{rf} energy towards \gls{rfid} tags, which reflect the \gls{rf} signal back to the transceiver, modulated with data. 
This approach is efficient for inventory management, asset tracking, and access control. While \gls{rfid} systems are useful, they require dedicated transceivers to enable backscattering for the \gls{rfid} tags. 
Further, the reader has to be in close proximity to \gls{rfid} tags for backscattering, which is a key limitation of RFID technology.
In contrast, \gls{aiot} leverages existing ambient \gls{rf} (cellular or TV) signals for operation, distinguishing it from traditional \gls{rfid} systems.
This adaptability positions \gls{aiot} as an exciting research area that aligns well with established communication infrastructures.

\section{Ambient IoT}\label{Ambient IoT}
\glsdesc{aiot} devices include both active devices with energy harvesting and passive devices with backscattering. 
These devices can be used in both monostatic and bistatic configurations depending on the requirements of the use cases.
\subsection{Comparison with existing IoT architectures}
\gls{aiot} devices are characterized by ultra-low complexity, compact size, limited capabilities, and an extended lifespan of approximately a decade. 
The distinctive features of \gls{aiot} devices further enhance their appeal for various applications.
These characteristics sharply contrast with existing \gls{lpwan} technologies like LoRaWAN, \gls{nbiot}, and \gls{emtc}, as seen in Table \ref{tab:comp_iot}. 
In essence, \gls{aiot} leverages backscattering ambient energy sources, targeting a different class of devices compared to existing \gls{iot} devices.

\begin{table}[!h]
\centering
\caption{Comparison of \gls{iot} network architectures}
\label{tab:comp_iot}
\footnotesize 
\begin{tabular}{p{2.2cm} p{0.8cm} p{2cm} p{1.5cm}}
\hline
\textbf{Network} & \textbf{Max} & \textbf{Coverage} & \textbf{Data Rates}\\
\textbf{Architecture} & \textbf{Power} & & \\
\hline
LoRaWAN & $25$mW & $10\text{-}15$km (rural) & $0.3\text{-}5.5$ kbps\\
\gls{nbiot} & $200$mW & $5\text{-}15$km (rural) & $250$ kbps \\
Active \gls{aiot} & $10$mW & $500$ m & $5$ kbps \\
Battery-free \gls{aiot}& $10\mu$W & $500$ m & $5$ kbps\\
\hline
\end{tabular}
\vspace{-0.3 cm}
\end{table}

\begin{table*}[!t]
\centering
\caption{Proposed specifications for \gls{aiot} devices}
\label{table:device_specifications}
\renewcommand{\arraystretch}{1.2} 
\begin{tabular}{|l|p{4.5cm}|p{3.5cm}|p{3.5cm}|}
\hline
\textbf{Device Type} & \textbf{Description} & \textbf{Power Consumption} & \textbf{Complexity} \\ \hline
\textbf{\Gls{bl}} & No energy source. Only backscatter communication. & $\leq 10\mu$W & Comparable to UHF RFID ISO18000-6C (EPC C1G2) \\ \hline
\textbf{\Gls{ba}} & Energy source for amplifying the backscattered signal. No independent signal generation. & Between \gls{bl} and BSA devices & Between \gls{bl} and BSA devices \\ \hline
\textbf{\Gls{bsa}} & Has an energy source. Can independently generate signal. & $\leq 10$mW & Much lower than NB-IoT devices \\ \hline
\end{tabular}
\end{table*}

\subsection{\gls{aiot} with \gls{3gpp}}
\gls{3gpp} has recently started discussions about \gls{aiot} devices and has included them in \glspl{tr} 38.848 \cite{38848} and 22.840 \cite{22840}. 
The specifications of \gls{aiot} in Rel-19 study item are summarized in \cite{nokia_ambient_iot}. 

The three types of \gls{aiot} devices are \glsdesc{bl} (BL), \glsdesc{ba} (BA), and \glsdesc{bsa} (BSA) devices. 
\gls{bl} devices have no energy source and no independent signal generation/amplification, relying solely on backscatter communication. \gls{ba} devices have an energy source for amplifying the backscattered signal but no independent signal generation. 
\gls{bsa} devices have an energy source and independent signal generation, using active \gls{rf} components for transmission. 
The major specifications of these devices are listed in Table \ref{table:device_specifications}.

\gls{aiot} devices are designed to support indoor environments with coverage ranging from 10 to 50 meters and outdoor environments from 50 to 500 meters. 
The data rate for uplink and downlink transmissions ranges from 0.1 kbps to 5 kbps. 
Each device can handle message sizes up to 1000 bits for reception and transmission. \gls{aiot} devices support up to 150 devices per 100 square meters indoors and up to 20 devices per 100 square meters outdoors.

\gls{3gpp} has defined the following network topologies for \gls{aiot} devices to connect to the network via \glspl{bs} and \glspl{ue}. In these topologies, the links may be unidirectional or bidirectional. The following main topologies are considered in recent \gls{3gpp} discussions:
\begin{enumerate}
    \item \gls{bs} $\leftrightarrow$ \gls{aiot} device: Direct, bidirectional communication.
    \item \gls{bs} $\leftrightarrow$ intermediate node $\leftrightarrow$ \gls{aiot} device: Bidirectional communication between \gls{aiot} device and the intermediate node (e.g., relay, \gls{iab} node, \gls{ue}, repeater).
\end{enumerate}
\section{Feasibility Study - Underground Backscattering}
In this section, we explore the use of backscattering technology for underground soil sensing in precision agriculture. 
This feasibility study examines the potential of backscattering communication for underground sensing, primarily focusing on \gls{aiot} devices, by leveraging UHF \gls{rfid} field trials. 
Given the similar device complexities between \gls{bl} \gls{aiot} and UHF \gls{rfid} devices, we conducted underground field trials using \gls{rfid} devices due to their commercial availability. Although our primary focus is on \gls{bl} \gls{aiot} devices, the comparable technology in \gls{rfid} allows us to draw relevant insights. 
We also perform a link budget analysis to compare the underground read ranges of \gls{bl} \gls{aiot} and \gls{rfid} technologies.
\begin{figure}[!t]
    \centering
    \includegraphics[width=0.85\linewidth]{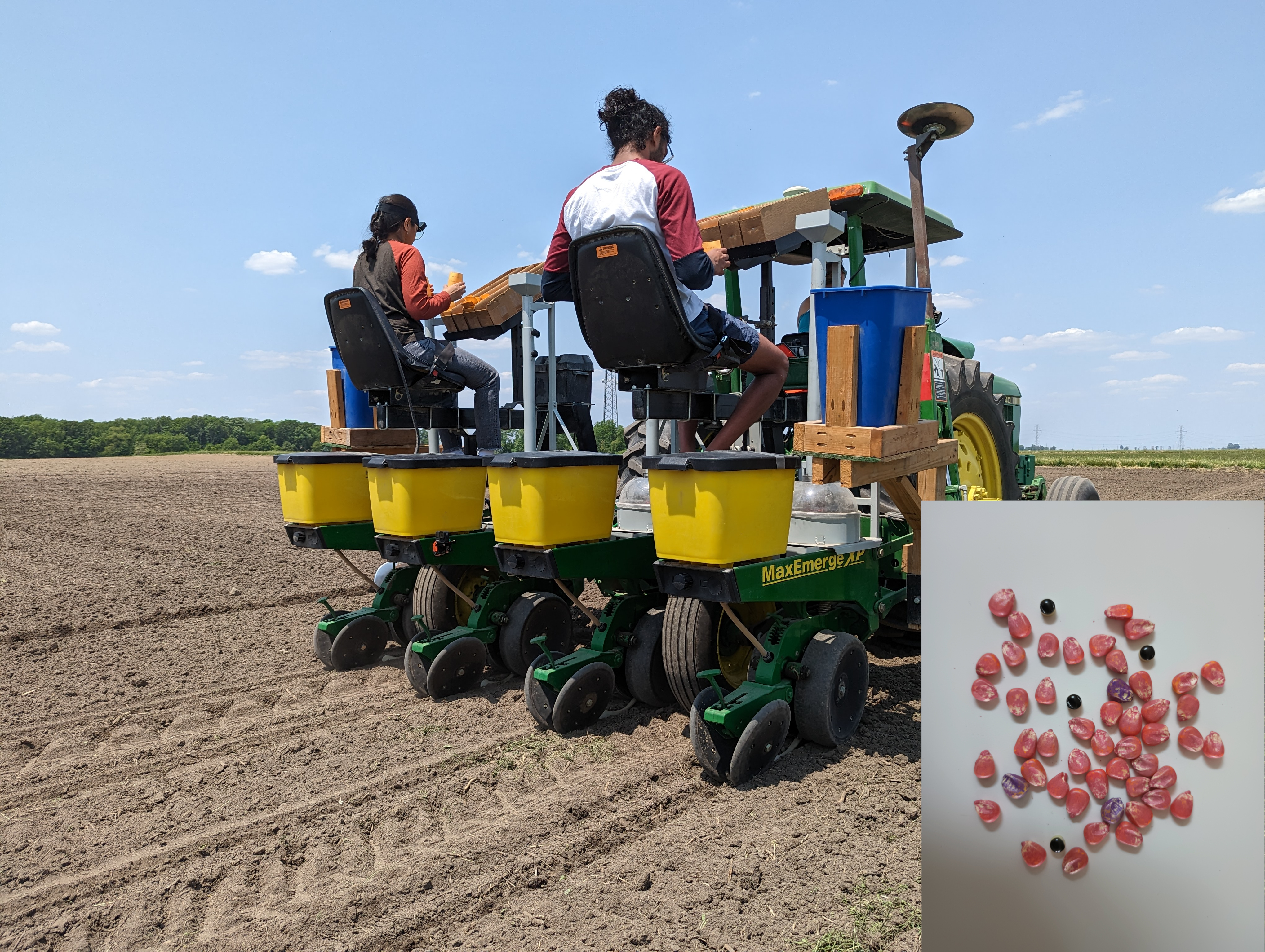}
    \caption{\gls{rfid} tags the size of corn seeds planted in the field with a conventional seed planter.}
    \vspace{-0.5cm}
    \label{fig:planter_seed}
\end{figure}
\subsection{Field Trials - Backscatter Communication for Underground Sensing}
To assess the feasibility of \gls{aiot} devices for underground sensing, a high-density deployment of sensors is necessary. 
Deploying a large number of sensors in agricultural settings presents several challenges. 
Currently, sensors are manually buried underground, a method that is labor-intensive and time-consuming, especially for large farms. 
Additionally, the deployment process must minimize disruption to existing crops. 
The high density required for effective monitoring necessitates innovative deployment practices to reduce labor and time.

In our field trials at Purdue University, we deployed \gls{rfid} tags underground using a conventional seed planter, as shown in Fig. \ref{fig:planter_seed}. 
Using \gls{rfid} tags comparable in size to corn seeds, we ensured compatibility with existing seed planters without modifications. 
This automated deployment method significantly reduces labor and time compared to manual deployment, offering benefits to farmers without adding overhead for sensor installation.
We planted $288$ \gls{rfid} tags in 12 rows of corn at a depth of $2.5$ cm underground and the number of tags were distributed based on the soil moisture levels.

The next challenge after deployment of these sensors is reading the data from these sensors.
Unlike traditional sensors that can automatically transmit and receive data, backscattering devices need excitation to activate these sensors for communication. 
Currently, \gls{rfid} devices are excited manually for devices deployed in a large area using \gls{rfid} readers. 
This is an arduous process to implement for a large farm to read the sensor data.
Unlike the wireless channel across air, the wireless channel across soil is harsh.
Apart from distance-based loss, there are losses from soil moisture, refraction etc. 
Thus, reading \gls{rfid} tags planted underground is much more difficult compared to terrestrial use cases.

\begin{figure}[!t]
    \centering
    \includegraphics[width=0.75\linewidth]{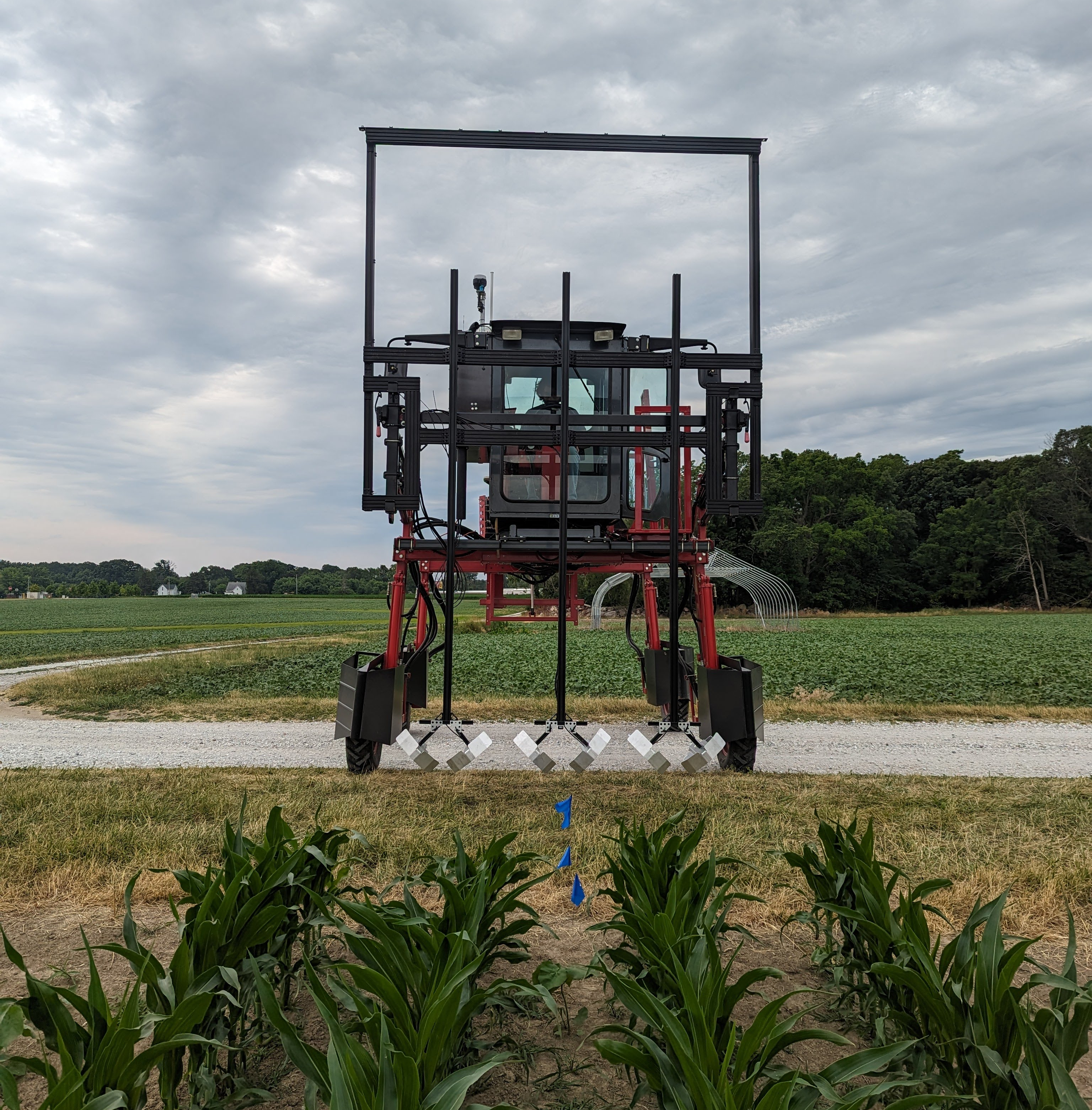}
    \caption{OATSMobile - communications platform enabling connected farms. OATSMobile is equipped with an \glsentrytext{rfid} reader and six antennas to read \glsentrytext{rfid} tags placed underground alongside corn plants in a field.}
    \vspace{-0.5cm}
    \label{fig:oats_mobile}
\end{figure}

In our field trials, we automated the reading process using OATSMobile, a communications platform enabling \textit{connected farms}. 
OATSMobile, a customized agricultural sprayer (Fig. \ref{fig:oats_mobile}), has six \gls{rfid} antennas mounted on the machine and an \gls{rfid} reader (Zebra FX9600) in its cabinet. 
The antennas were strategically placed to optimize tag detection, with two antennas directed towards each row of corn where the \gls{rfid} tags were planted.
As OATSMobile moves through the farm, it positions the antennas between the rows of corn. 
In a single pass, it covers four rows, reducing obstruction between antennas and plants. 
The experiment was conducted at different corn growth stages to observe seasonal effects. 
Out of 288 planted \gls{rfid} tags, 152 unique tags were successfully read, resulting in a $53.9\%$ reception success rate. 
The success rate was influenced by factors such as soil moisture, corn canopy growth, and tag depth. 
The link-budget analysis in the following section corroborates the impact of soil moisture on the read range of the backscattering devices.

\subsection{Underground Backscattering Link Budget Analysis}
We conduct a link budget analysis for both excitation (downlink) and backscattered (uplink) connections of \gls{aiot} and \gls{rfid} devices placed underground to understand our field trials. 
This evaluation allows us to compare the read ranges of \gls{aiot} devices with \gls{rfid} devices. 
In our analysis, we considered the transmitter and reader above the ground, and the tag, which is either underground.


For tags placed underground, the analysis is complex due to soil channel characteristics. 
The downlink is an \gls{ag2ug} link and the uplink is an \gls{ug2ag} link. 
The received tag power is $P_{rx,tag} = P_T G_T G_{tag} L_{AG-UG}(d_1)$
where $P_T$ is the transmitter power, $G_T$ is the transmit antenna gain, $G_{tag}$ is the tag antenna gain, $L_{AG-UG}(d)$ is the \gls{ag2ug} path loss and $d_1$ is the distance between the transmitter and the tag \cite{backscatter_link_budget}. 
Similarly, the backscattered power received in the reader is $P_{rx,read} = P_{rx,tag} G_{tag} G_R M L_{UG-AG}(d_2)$ where $G_R$ is the reader antenna gain, $M$ is the backscatter modulation factor, and $L_{UG-AG}(d)$ is the \gls{ug2ag} path loss. 
Both $L_{AG-UG}(d)$ and $L_{UG-AG}(d)$ include aboveground path loss, underground path loss, and refractive loss components \cite{underground_channel_model}. 
The underground path loss depends on factors such as volumetric water content and soil clay fraction, which alter soil permittivity.

In monostatic cases, $d_1 = d_2$ and $G_T = G_R$ as the exciter and the reader are the same device. 
In bistatic cases, $d_1 \neq d_2$, and $G_T = G_R$ if the transmitter and reader use the same antenna gain, otherwise $G_T \neq G_R$. 
For successful communication, the following conditions must be met:
\begin{itemize}
    \item Received tag power $P_{rx,tag}$ should be exceed the tag activation threshold $P_{thr}$. Only if this condition is met, the tag wakes up. The distance $d_{act}$ at which $P_{rx,tag} = P_{thr}$ is called the DL distance or activation distance.
    \item Received backscattered power $P_{rx,read}$ should exceed the reader sensitivity $S$. The distance $d_{read}$ at which $P_{rx,read} = S$ is named the UL distance or read distance.
\end{itemize}

\begin{figure}[!t]
    \centering
    \includegraphics[width=\linewidth]{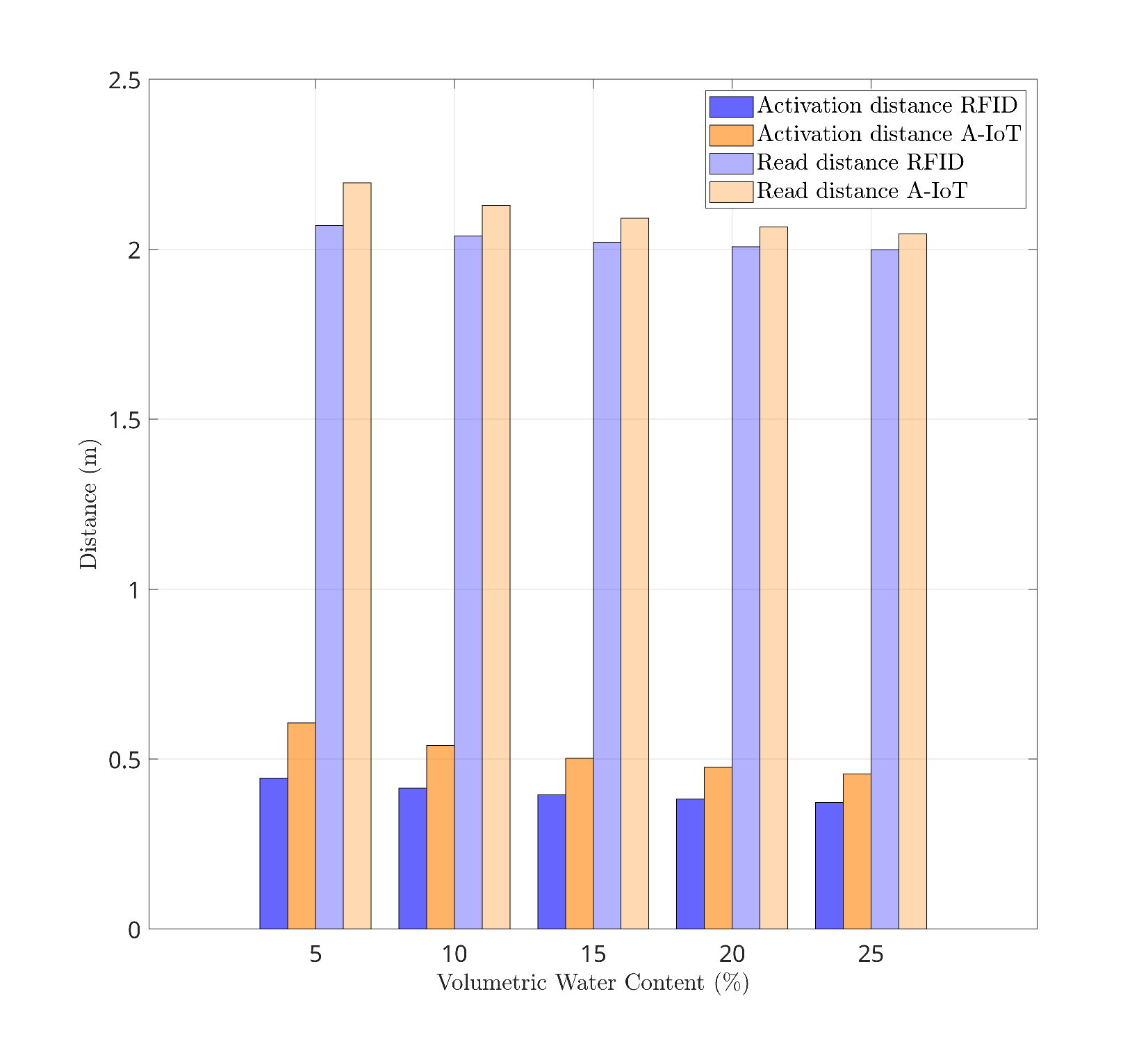}
    \caption{Activation distance and read distance of \gls{aiot} and \gls{rfid} devices placed underground in both monostatic and bistatic configurations. In a monostatic configuration, $d_{act} = d_{read}$ while in a bistatic configuration, $d_{act} \neq d_{read}$.}
    \vspace{-0.5cm}
    \label{fig:undergroundTags}
\end{figure}

The link budget analysis parameters for \gls{rfid} and \gls{aiot} systems are as follows: the transmitted power ($P_T$) is $30$ dBm for \gls{rfid} and $24$ dBm for \gls{aiot}. 
The tag threshold power ($P_{\text{thr}}$) is $-10$ dBm for \gls{rfid} and $-25$ dBm for \gls{aiot}, with receiver sensitivities ($S$) of $-75$ dBm and $-100$ dBm, respectively. 
The modulation factors ($M$) are $0.33$ for OOK in \gls{rfid} and $0.25$ for \gls{aiot}. 
Both systems share a path loss exponent ($\gamma$) of $3$, and antenna gains ($G_T = G_R$) of $6$ dBi, with a tag gain ($G_{\text{tag}}$) of $-1$ dB. 
The transmitter is placed $0.3$ meters aboveground, and the volumetric water content (VWC) ranges from $5\%$ to $25\%$.
Based on these parameters, we calculated activation and read distances for both \gls{aiot} and \gls{rfid} devices and are illustrated in Fig. \ref{fig:undergroundTags}.

Our results demonstrate the superior read range capabilities of \gls{aiot} devices compared to \gls{rfid} for tags placed underground in monostatic and bistatic configurations. \gls{aiot} tags exhibit a higher activation and reading distance than \gls{rfid} tags, although the activation distance for underground tags is reduced due to higher soil path loss. Additionally, the data shows a decrease in $d_{act}$ and $d_{read}$ with increasing \gls{vwc}, indicating higher path loss in wetter soil. These findings suggest that \gls{aiot} technology, with its lower activation threshold and higher sensitivity, is better suited for robust and reliable subterranean soil sensing applications, enhancing underground communication and precision agriculture techniques.

\section{Challenges and Research Directions}\label{challenges}

\subsection{Challenges}
\glsdesc{aiot} faces challenges due to lack of coordination between transmitter and receiver, making it difficult to know the \gls{csi}. 
The ambient \gls{rf} signal can interfere with the receiver, complicating the detection process. 
While advanced machine learning techniques can improve detection and estimation, the simplicity and battery-free nature of \gls{aiot} devices mean that existing wireless algorithms cannot be applied directly. Therefore, new techniques are needed to overcome these challenges.

\subsubsection{Energy Harvesting}
Despite the simplicity, backscattering has many challenges in providing reliable communication for \gls{aiot} devices. 
Some of the key research problems in backscattering comprise energy efficiency, communication range, network responsiveness etc. 
The battery-free and ultra-low power operation of backscattering makes it a good technique for \gls{aiot} devices. 
The ambient \gls{rf} signals are used for backscattering and typically \gls{rf} harvesting efficiency is as low as 18.2\% \cite{xu2018practical}.
Research could be directed on improving the energy efficiency of backscattering systems for \gls{rf} signals.

\subsubsection{Backscatter Signal Detection/Estimation}
 Unlike traditional communication systems, in a backscatter scheme, transmitting bits `1' and `0' correspond to whether the device is backscattering or not.
Since \gls{aiot} devices merely backscatter existing \gls{rf} signals, it becomes challenging for the reader to detect and estimate the data as it cannot estimate the \gls{csi} without pilot symbols \cite{wang2016ambient}. 
Implementing \gls{ml}-based techniques is also challenging because the acceptable error rates for detection are much lower than those in traditional \gls{ml} applications \cite{liu2020deep}. 
Therefore, further research is needed in the detection and estimation of ambient backscattering signals, given the lack of coordination between the device and the reader.

\subsubsection{Security}
Security is one of the major aspects in any network. In \gls{aiot} systems, the devices are limited in complexity and therefore cannot employ the complex techniques used in 5G-NR.
Also, security in \gls{aiot} systems should be focused in physical layer as devices have minimal upper layer components unlike traditional systems. These devices are very easy to eavesdrop on as they are very similar to \gls{rfid} systems. 
We need to ensure authentication and confidentiality over physical layer security.
There is a need to develop new security designs and protocols that work in low-complexity devices such as \gls{aiot} devices. 

\subsubsection{Access for Stateless Devices}
The challenge lies in communicating with devices that have low or no stored energy and cannot always respond to network paging signals. 
Current cellular protocols rely on paging mechanisms with \gls{rrc} states, assuming devices are always reachable and can initiate paging. However, \gls{aiot} devices lack defined \gls{rrc} states, making these assumptions invalid and necessitating new access mechanisms.

\subsection{Research Directions}
\gls{aiot} presents unique challenges, including lack of coordination between devices and the reader. This section explores potential avenues to improve connectivity and energy efficiency of \gls{aiot} devices.


\subsubsection{MIMO Backscatter Communications}
\Gls{mimo} has been a key technology, improving the capacity and reliability of wireless networks. 
However, its use in backscatter communications has been limited, despite its potential to enhance range, capacity, and reliability. 
Multi-antenna readers can increase the read range through transmit and receive beamforming, which requires estimating the \gls{mimo} backscatter channel. 
Employing multiple antennas on tags is challenging due to their passive nature.

Several open research problems must be addressed to facilitate \gls{mimo} techniques such as beamforming. 
First, backscatter channel estimation is critical. 
While acquiring \gls{csi} has been widely studied in prior \gls{mimo} work, backscatter channel estimation remains unexplored due to its unique cascaded structure with forward and backward channels, complicating the signal model. 
Initial work for the monostatic scenario exists \cite{mishra2019multi}, but bistatic/ambient scenarios remain unexplored. 
Next, space-time codes can be promising for multi-antenna tag precoding. 
Passive tags cannot adapt to the channel due to their nature, but space-time codes (e.g. Alamouti) can achieve diversity gain without explicit \gls{csi}. 
A novel space-time code proposed in \cite{luan2021better} outperforms the Alamouti code for multi-antenna tags.


\subsubsection{Multiple Access / Random Access}
\gls{aiot} devices have ultra-low complexity and a large device density. In order to support the large \gls{aiot} device density concurrently, we need state-of-the-art multiple access techniques. Since \gls{aiot} devices have high device density, random access techniques are employed to support high device density. For UHF \gls{rfid} devices, schemes such as slotted ALOHA, Q-protocol are used for random access. Since \gls{bl} is of the same complexity as UHF \gls{rfid}, \gls{aiot} devices could use similar random access techniques. There are research opportunities in finding better multiple access techniques for all devices types in \gls{aiot}.

\subsubsection{Advanced Modulation Techniques}
In backscattering communication, the load impedance of the \gls{aiot} device is varied to modulate the reflected signal to the reader. Currently, simpler modulation schemes such as \gls{ask}, \gls{psk} are used in backscattering. For larger data rates, we need larger number of load impedance states in order to support higher order modulation schemes with low complexity. 


\subsubsection{Security}
\gls{aiot} security should guarantee authentication and confidentiality for every device in the network. Unlike traditional wireless systems, \gls{aiot} devices have lower complexity. Hence, we should focus on physical layer security and algorithms beyond 5G-NR \cite{ruzomberka_security}. This is achievable via lightweight authentication protocols such as hash functions etc. 

%
\subsubsection{Positioning}
Positioning \gls{aiot} devices is crucial for various precision agriculture applications, enabling the tracking of farm assets, livestock, and machinery. 
However, passive \gls{aiot} devices pose challenges since they cannot process received signals, making downlink-based positioning difficult. 
Instead, localization should be done by the reader, though large inter-site distances in rural networks can hinder positioning due to low signal strength. 
Moving nodes like tractors and drones can read backscattered signals and locate \gls{aiot} devices in a bistatic/multi-static manner, with location data reported to \gls{lmf}. Additionally, \gls{aiot} devices can serve as anchors to help localize other \gls{3gpp} nodes, enhancing positioning through triangulation without constructing new \glspl{pru}. 
This approach is particularly beneficial in rural areas with low \gls{3gpp} node density.

\section{Conclusion}\label{conclusion}
This work provides an overview of \glsdesc{aiot}, a simple, low-cost and battery-free technology with significant potential for precision agriculture. This article explored various use cases of \gls{aiot} in this field, highlighting its advantages over \gls{rfid}, particularly for underground sensing. Despite promising prospects, several challenges need to be addressed for \gls{aiot} to become a reality. These include the need for multiple access, positioning and efficient multiple access. By exploring these research directions, we can harness the full potential of \gls{aiot} in precision agriculture. This article serves as a comprehensive summary of \gls{aiot} and its promising applications in advancing agricultural practices.

\bibliographystyle{IEEEtran}
\bibliography{references}

\begin{IEEEbiographynophoto}{Ashwin Natraj Arun} is currently working toward a Ph.D. degree in Electrical and Computer Engineering at Purdue University, West Lafayette, IN, USA. 
His research interests include signal processing for wireless communications and information theory.
\end{IEEEbiographynophoto}

\begin{IEEEbiographynophoto}{Byunghyun Lee}
received his B.S. and M.S. degree in electrical engineering from Yonsei University, Seoul, South Korea, in 2017 and 2019, respectively. 
He is currently pursuing his Ph.D. in electrical and computer engineering at Purdue University, West Lafayette, IN, USA. 
His research interests include beamforming and estimation/tracking techniques for integrated sensing and communication systems.
\end{IEEEbiographynophoto}

\begin{IEEEbiographynophoto}{Fabio A. Castiblanco} is currently working toward a Ph.D. degree in electrical and computer engineering with Purdue University, West Lafayette, IN, USA. 
His research interests include embedded systems, computer networking, and wireless communications. 
Castiblanco received a B.S. degree in electronics engineering from the Universidad Nacional de Colombia (UNAL), Bogota, Colombia. 
\end{IEEEbiographynophoto}

\begin{IEEEbiographynophoto}{Dennis R. Buckmaster} received the B.S. from Purdue University and  M.S. and Ph.D. degrees in Agricultural Engineering from Michigan State University in 1984, 1986 and 1989, respectively. 
He is a Professor of Purdue ABE. 
His current research and teaching interests in agricultural systems management are in all aspects of the data pipeline of digital agriculture including sensors, data flow, analytics, biophysical modeling for decision making, and automation; his application focus is cropping systems and livestock feeding systems. 
He is an award winning teacher and mentor.
\end{IEEEbiographynophoto}

\begin{IEEEbiographynophoto}{Chih-Chun Wang} is a Professor of Purdue ECE. 
His current research interests are in the latency minimization of 6G-and-beyond wireless networks, information and coding theory for ultra-low latency communications. 
He received the NSF CAREER Award in 2009. He is an IEEE Fellow.
\end{IEEEbiographynophoto}

\begin{IEEEbiographynophoto}{David J. Love} is the Nick Trbovich Professor of Electrical and Computer Engineering at Purdue University.  He is a Fellow of the IEEE, American Association for the Advancement of Science (AAAS), and National Academy of Inventors (NAI).
    
\end{IEEEbiographynophoto}

\begin{IEEEbiographynophoto}{James V. Krogmeier} received the BSEE degree from the University of Colorado and the MS and Ph.D. degrees from the University of Illinois. He is currently Professor of Electrical and Computer Engineering at Purdue and his research interests are in statistical signal processing.
    
\end{IEEEbiographynophoto}

\begin{IEEEbiographynophoto}{M. Majid Butt} is a senior staff research scientist at Nokia, USA.  
He has authored more than 80 peer-reviewed articles and contributed to more than 100 filed/granted patents in cellular networks domain. 
He serves as an associate editor for IEEE Open Journal of the Communication Society and IEEE Open Journal of Vehicular Technology. 
He is a senior member of IEEE and IEEE COMSOC distinguished lecturer.
\end{IEEEbiographynophoto}

\begin{IEEEbiographynophoto}{Amitabha (Amitava) Ghosh} (F’15)  is a Nokia Fellow and works at Nokia Standards and Strategy. 
He joined Motorola in 1990 after receiving his Ph.D in Electrical Engineering from Southern Methodist University, Dallas. 
Since joining Motorola he worked on multiple wireless technologies starting from IS-95 to 3GPP New Radio (NR). 
He has more than 65 issued patents, has written multiple books and book chapters and has authored numerous technical papers. 
He is also the chair of the NextG alliance, National Roadmap Working Group.
\end{IEEEbiographynophoto}

\end{document}